\documentclass[11pt,catalan]{article}
\usepackage{graphicx}
\usepackage{amsmath}
\pdfoutput=1
\usepackage{amsfonts}
\usepackage{amssymb}
\usepackage{bm}

\usepackage[english]{babel}
\usepackage[latin1]{inputenc}
\usepackage{times}
\usepackage[T1]{fontenc}

\newcommand{\D}{{\rm d}}

\newcommand{\imap}{{\mathbb{I}}{\rm m}}
\newcommand{\realp}{{\mathbb{R}}{\rm e}}

\begin{document}
\title{Energy-momentum tensor for the electromagnetic field\\in a dispersive medium as an application of Noether theorem}
\author{Carlos Heredia\thanks{e-mail address: carlosherediapimienta@gmail.com}\,\,\, and Josep Llosa\thanks{e-mail address: pitu.llosa@ub.edu}\\
Facultat de F\'{\i}sica (FQA and ICC) \\ Universitat de Barcelona, Diagonal 645, 08028 Barcelona,
Catalonia, Spain }
\maketitle

\begin{abstract}
On the basis of a non-local Lagrangian for Maxwell equations in a dispersive medium, the energy-momentum tensor of the field is derived. 
We obtain the Field equations through variational methods and an extension of Noether theorem for a non-local Lagrangian is obtained as well.
The electromagnetic energy-momentum tensor obtained in the general context is then specialized to the case of a field with slowly varying amplitude on a rapidly oscillating carrier. 
\\[1ex]
\noindent
\end{abstract}

\section{Introduction }
The electromagnetic field produced by a distribution of free charge and current in a material medium is ruled by Maxwell equations
\begin{equation}  \label{e1}
\nabla\cdot \mathbf{B} = 0  \,, \qquad \qquad \nabla\times \mathbf{E} + \partial_t \mathbf{B} = 0 
\end{equation}
and
\begin{equation}  \label{e2}
\nabla\cdot \mathbf{D} = 4 \pi\,\rho  \,, \qquad \qquad \nabla\times \mathbf{H} - \partial_t \mathbf{D} = 4 \pi\,\mathbf{j} 
\end{equation}
where $\mathbf{E}$ and $\mathbf{H}$ respectively are the electric and magnetic fields, $\mathbf{D}$ is the electric displacement, $\mathbf{B}$ is magnetic induction and $\rho$ and $\mathbf{j}$ are the free charge and current densities (unrationalized Gaussian units with $c=1$ have been assumed). 
$\mathbf{E}$ and $\mathbf{B}$ are the physical magnitudes as they manifest in the Lorentz force on a test charge.
The above system  (\ref{e1}-\ref{e2}) does not determine $\mathbf{E}$ and $\mathbf{B}$ if only the distributions of charge and current are known, because the number of unknowns largely exceeds the number of equations. 
This hindrance is circumvented by specifying the nature of the material medium, i. e. giving the constitutive equations, a set of phenomenological relations connecting $\mathbf{D}$ and $\mathbf{H}$ with the physical variables $\mathbf{E}$ and $\mathbf{B}$.

For isotropic non-dispersive linear media the constitutive equations are
$\,\mathbf{D} = \varepsilon \mathbf{E}\,$ 
and $\,\mathbf{H}=\mu^{-1}\,\mathbf{B} \,$, 
where $\varepsilon$ and $\mu$ respectively are the dielectric and magnetic constants. This is the case most considered in textbooks \cite{Jackson} and also in  the seminal paper \cite{Minkowski1908} where Minkowski set up the relativistic electrodynamics in material media and, particularly, derived his (non-simetric) energy-momentum tensor for the electromagnetic field.
Vacuum is a particular case with $\,\varepsilon= \varepsilon_0\,$ and $\,\mu= \mu_0\,$.

However, in natural media $\varepsilon$ and $\mu $ are not constant and  generally depend on frequency and wavelength. We then speak of dispersive media. For a plane monochromatic wave, $\mathbf{E}(\mathbf{q},\omega) \,e^{i(\mathbf{q}\cdot\mathbf{x}-\omega t)}\,$ 
and $\,\mathbf{B}(\mathbf{q},\omega) \,e^{i(\mathbf{q}\cdot\mathbf{x}-\omega t)}\,$, 
the response of the medium is a displacement $\mathbf{D}(\mathbf{q},\omega) \,e^{i(\mathbf{q}\cdot\mathbf{x}-\omega t)}\,$ and a magnetic field $\mathbf{H}(\mathbf{q},\omega) \,e^{i(\mathbf{q}\cdot\mathbf{x}-\omega t)}\,$, with
\begin{equation}  \label{e4}
\mathbf{D}(\mathbf{q},\omega) = \varepsilon(\mathbf{q},\omega)   \, \mathbf{E}(\mathbf{q},\omega)
\,, \qquad \qquad \mathbf{H}(\mathbf{q},\omega) = \mu^{-1}(\mathbf{q},\omega)   \, \mathbf{B}(\mathbf{q},\omega)
\end{equation}
By Fourier transform we express a general electromagnetic field, $\mathbf{E}(\mathbf{x},t)$ and $\mathbf{B}(\mathbf{x},t)$ as a superposition of plane monochromatic waves, each of them producing an electric displacement and magnetic field like (\ref{e4}). By the convolution theorem \cite{mates1}, the superposition of all of them results in
\begin{equation}  \label{e5}
\mathbf{D}  = (2\pi)^{-2}\,\tilde\varepsilon \ast \mathbf{E} \,, \qquad \qquad \mathbf{H}  = (2\pi)^{-2}\,\widetilde{\mu^{-1}} \ast \mathbf{B}
\end{equation}
where $\tilde\varepsilon$ and $\widetilde{\mu^{-1}}$ are connected with the Fourier transforms of $\varepsilon$ and $\mu^{-1}$, that is
\begin{equation}  \label{e5a}
\tilde\varepsilon(y) = (2\pi)^{-2} \int \D^4 k\,\varepsilon(k) e^{ik_b y^b} \,, \qquad \qquad \varepsilon(k) = (2\pi)^{-2} \int \D^4 y\,\tilde\varepsilon(y)  e^{-ik_b y^b} 
\end{equation}
where $\,  k^b = (\mathbf{q} ,\omega)  \,$.

Complemented with these constitutive relations, the system (\ref{e1}-\ref{e2}) allows to determine the electromagnetic fields, $\mathbf{E} $ and $\mathbf{B}$  ---hence the Lorentz force on any test charge--- provided that we know: (a) the distribution of free charge and current, (b) the nature of the medium specified by the dielectric and  magnetic function,   $\varepsilon(\mathbf{q},\omega) $ and $\mu(\mathbf{q},\omega) $, and (c) the suitable boundary conditions for such a partial differential system.

Without leaving the mathematical framework described so far we can modify the variables of our problem and their interpretation by replacing $\mathbf{D} $ and $\mathbf{H}$ by some new variables that describe the collective behavior of the elementary charges in the material medium, namely the polarization and magnetization densities \cite{Jackson1}
\begin{equation}  \label{e6}
\mathbf{D}  = \mathbf{E} + 4 \pi\,  \mathbf{P} \,, \qquad \qquad \mathbf{H}  =  \mathbf{B} - 4\pi\,\mathbf{M}
\end{equation}
Using this, the inhomogeneous pair of Maxwell equations (\ref{e2}) becomes
\begin{equation}  \label{e7}
\nabla\cdot \mathbf{E} = 4 \pi\,(\rho + \rho_b) \,,\qquad \qquad \nabla\times \mathbf{B}- \partial_t \mathbf{E} = 4 \pi\, (\mathbf{j} + \mathbf{j}_b)
\end{equation}
where
\begin{equation}  \label{e8}
\rho_b = - \nabla\cdot \mathbf{P}  \qquad {\rm and} \qquad \mathbf{j}_b = \nabla \times \mathbf{M} + \partial_t \mathbf{P}
\end{equation}
respectively, the density of bound charge and bound current (in contrast with the free charge distributions $\rho$ and $\mathbf{j}$).

For an isotropic dispersive medium the definitions (\ref{e6}) and the equations  (\ref{e5}) imply that
\begin{equation}  \label{e9}
\mathbf{P}  = \tilde\chi_e \ast \mathbf{E} \,, \qquad \qquad \mathbf{M}  = \tilde\chi_m \ast \mathbf{B}
\end{equation}
where 
$$ \tilde\chi_e = \frac1{4\pi}\,\left[\tilde\varepsilon - \delta(\mathbf{x})\,\delta(t) \right]\qquad {\rm and} \qquad 
\tilde\chi_m = \frac1{4\pi}\,\left[\delta(\mathbf{x})\,\delta(t) - \widetilde{\mu^{-1}}- \mu_0^{-1}\right]
$$
are the electric and magnetic  susceptibilities of the medium.

As Maxwell equations (\ref{e1}-\ref{e2}) and the constitutive relations (\ref{e5}) are linear, we can split the  electromagnetic fields as
$$  \mathbf{E} = \mathbf{E}_0 + \mathbf{E}_{\rm ind} \,, \qquad  \mathbf{B} = \mathbf{B}_0 + \mathbf{B}_{\rm ind}  \,, $$
i. e. the sum of two contributions: whereas $\mathbf{E}_0$ and $\mathbf{B}_0$ are the solution of Maxwell equations in vacuum for the distribution of free charges and currents, $\mathbf{E}_{\rm ind}$ and $\mathbf{B}_{\rm ind}$ are the solution of Maxwell equations in vacuum for the distribution of bound charges and currents. The interpretation that follows is: free charges and currents produce the electromagnetic field,  $\mathbf{E}_0$ and $\mathbf{B}_0$, which polarizes the medium. This polarization implies a distribution of bound charges and currents which in turn produce the induced electromagnetic field, $\mathbf{E}_{\rm ind}$ and $\mathbf{B}_{\rm ind}$. The latter is physically indistinguishable of the primary field and only the total field, $\mathbf{E}$ and $\mathbf{B}$, manifests trough the total Lorentz force on a test charge.

As far as the resolution of Maxwell equations is concerned, this second view is not practical at all, however it will be helpful and illuminating to understand the exchange of energy and momentum between the electromagnetic field and the electric charges, either free or bound.
 
Poynting theorem \cite{JacksonPoynting} is about the energy exchange between the free charges and the electromagnetic field and it holds for non-dispersive media only. Its derivation follows from the scalar product of $\mathbf{E}$ and the second equation (\ref{e2}). Then, a vector identity is invoked with the need of assuming that the dielectric and magnetic functions are constant. The quantity  $\,\displaystyle{\mathcal{U} = \frac1{8\pi}\,\left[\mathbf{E}\cdot \mathbf{D} + \mathbf{B}\cdot \mathbf{H} \right] }\, $ and the Poynting vector $\,\mathbf{S}\,$ are respectively taken as the energy density of the electromagnetic field  and the current density of energy. The main idea at the back of the theorem is that the increase of the field energy and the kinetic energy of free electric charges in a region in space is due to the energy flowing through its boundary. To obtain the linear momentum balance, one can do similarly \cite{JacksonPoynting} although the procedure is much more elaborated. 

However the spacetime formalism introduced by Minkowski \cite{Minkowski1908} is largely simpler. It treats the energy and momentum exchanges on the same foot and proceeds similarly as in the proof of Poynting theorem, combining some differential tensor identity and Maxwell equations.

As previously mentioned, Poynting theorem does not hold for dispersive media. In our view this is due to the fact that the energy-momentum balance should also include the energy and momentum stored in the ``bound charges'' which will depend, in the end, on the polarization and magnetization densities, $\mathbf{P}$ and $\mathbf{M}$, and maybe on their derivatives as well.

The plan of the paper is as follows. In Section 2 we outline the electrodynamics of Minkowski \cite{Minkowski1908} for media with constant dielectric and magnetic functions. Furthermore, we use an action principle and a Lagrangian for the electromagnetic field in the medium that, applying Noether theorem, yields a canonical conserved energy-momentum tensor. Finally, we find the associated Belinfante-Rosenfeld tensor ---see for instance \cite{Belinfante} and the outline in the Appendix. The latter coincides with the non-symmetric energy-momentum tensor proposed by Minkowski by merely elaborating from the field equations. 

In Section 3 we generalize Minkowski electrodynamics to dispersive media. This leads to a non-local Lagrangian density, i. e. it contains a convolution product whose value at the point $x$ depends on the values of the field at any point in spacetime. We then derive the field equations and apply Noether theorem to obtain a conserved energy-momentum tensor. 
As we are aware that non-local Lagrangians are seldom found in textbooks, we devote the Appendix to outline the derivation of the field equations and the generalisation of Noether theorem for such Lagrangians.

\section{Outline of Minkowski electrodynamics (non-dispersive media)  \label{S2}}
This is a version of Minkowski's proposal in a notation closer to that used in present time textbooks \cite{Dixon}. 
In the spacetime formalism the coordinates in an inertial reference system are denoted as 
\begin{equation}  \label{e10a}
x^1=x\,, \qquad \quad x^2 = y\,, \qquad \quad x^3 = z \,, \qquad \quad x^4 = t \,;
\end{equation}
the electromagnetic field is represented by Faraday tensor
\begin{equation}  \label{e10}
F_{ab}= - F_{ba} \,, \qquad \quad 
  F_{12} = B_3 \,, \qquad \quad  F_{23} = B_1\,, \qquad \quad   F_{31} = B_2 \,, \qquad \quad F_{i4} = E_i
\end{equation}
with $a,b = 1 \ldots 4\,$ and similarly the electric displacement $\mathbf{D}$  and the magnetic field $\mathbf{H}$ are represented by the {\em displacement tensor}
\begin{equation}  \label{e11}
H^{ab}= - H^{ba} \,, \qquad \quad  H^{12} = H_3 \,, \qquad \quad  H^{23} = H_1\,, \qquad \quad   H^{31} = H_2 
\,, \qquad \quad H^{i4} = - D_i
\end{equation}
Adopting the notation
$$ \partial_a = \frac{\partial\;\;}{\partial x^a}\,, \qquad \quad \nabla = (\partial_1, \,\partial_2, \,\partial_3)  \,, \qquad \quad \partial_4 = \frac{\partial\;\;}{\partial t}  \,, $$
Maxwell equations (\ref{e1}-\ref{e2}) can be respectively written as
\begin{equation}  \label{e12a}
 \partial_a F_{bc} + \partial_b F_{ca} + \partial_c F_{ab} = 0   \qquad\quad {\rm and} \qquad\quad
 \partial_b H^{ab}  = J^a
\end{equation}
where 
$$ J^1 = j_x \,, \qquad \quad  J^2 = j_y \,, \qquad \quad  J^3 = j_z \,, \qquad \quad  J^4 = \rho $$
is a 4-vector made of the free charge and current densities.

This arrangement is specially suited to deal with coordinate transformations connecting two inertial systems. Indeed, given two systems of coordinates $\, (x^1,\, \, x^2 ,\, x^3,\, x^4)\, $ and $\, (x^{\prime 1},\, \, x^{\prime 2} ,\, x^{\prime 3},\, x^{\prime 4})\, $ connected by a  Poincar\'e transformation $\, x^{\prime a} = \Lambda^a_{\;b} x^b + s^b \,$ 
(where $\,\Lambda^a_{\;b}\,$ is a Lorentz matrix and $s^b$ four constants), then we have that
\begin{equation}   \label{e12d}
 F^\prime_{ab} = \Lambda^c_{\;a} \Lambda^d_{\;b} F_{cd} \,,\qquad \quad 
 H^{\prime\,ab} = \Lambda^a_{\;c} \Lambda^b_{\;d} H^{cd}  \qquad \quad {\rm and} \qquad \quad  
J^{\prime a} = \Lambda^{a}_{\; b} J^b 
\end{equation}
A relevant role is reserved to the Minkowski metric 
$$ \eta_{ab} = {\rm diag}(1 \,1 \, 1 \, -1 ) \,,$$
and its inverse $\,\eta^{ab} = {\rm diag}(1 \,1 \, 1 \, -1 ) \,$, such that $\quad \eta^{ac} \eta_{cb} = \delta^a_b \,$ (the Einstein convention of summation over repeated indices is adopted).

The same tensor symbol may occur sometimes with the indices in the lower position (covariant) or in the upper position (contravariant). The relation between them is, e. g. 
$$ F_{\; b}^a = \eta^{ac} F_{cb}  \qquad \quad {\rm or} \qquad \quad F^{ab} = \eta^{ac} \eta^{bd} F_{cd} $$

The first Maxwell equation (\ref{e12a}) means that the Faraday tensor can be derived from a 4-potencial \cite{Dixon}
\begin{equation}   \label{e12c}
\mbox{it exists} \quad  A_a \quad\mbox{ such that}\qquad   F_{ab} = \partial_a A_b - \partial_b A_a
\end{equation}

As pointed out before, the system (\ref{e12a}) must be  complemented with a set of constitutive relations that, for a non-dispersive homogeneous and isotropic medium, are
$$ \mathbf{D} = \varepsilon \, \mathbf{E} \,, \qquad \qquad \mathbf{H} = \mu^{-1} \, \mathbf{B} $$
However, as the Poincar\'e transformation (\ref{e12d}) entangles the electric and magnetic parts, this simple form cannot be valid in all inertial frames but only in the {\em proper inertial frame}, i. e. that respect to which the medium is at rest.

\paragraph{Electric and magnetic fields.} An inertial reference system is characteritzed by its proper velocity with respect to the laboratory frame; this is a  4-vector $\,u^a = (\gamma\,\mathbf{u}, \gamma)\,$, where $\,\gamma = (1-v^2)^{-1/2}\,$ and $\mathbf{v}$ is 
the standard 3-velocity. It is a timelike unit vector, that is
$$ u^a u_a = u^a u^b \eta_{ab} = -1 $$.

Given any skewsymmetric tensor as $F_{ab}$, they exist $\,E_b$ and $\, B^d\,$ such that
\begin{equation}   \label{e13}
    F_{ab} = 2\,u_{[a} E_{b]} + \hat{F}_{ab}  \,, \qquad \qquad    \hat{F}_{ab}= \varepsilon_{abcd} u^c B^d \,, \qquad \qquad  u^a E_a = u_d B^d = 0 
\end{equation}
where the square bracket means antisymmetrization and $\,\varepsilon_{abcd} \,$ is the totally skewsymmetric Levi-Civita symbol in 4 dimensions:
$$ \varepsilon_{abcd} = \left\{ \begin{array}{cl}
                              -1 & \mbox{if $abcd$ is an even permutation of 1234} \\
                              1 & \mbox{if $abcd$ is an odd permutation of 1234} \\
                              0 & \mbox{if there is some repeated index} 
															\end{array}  \right.   $$
It can be easily checked that 
\begin{equation}   \label{e13a}
   E_a = F_{ab} u^b  \qquad {\rm and} \qquad B^d = \frac12\, \varepsilon^{cdab} u_c F_{ab}
\end{equation}
where we have used that $u^a$ is a unit vector and that 
$$ \varepsilon^{abcd}\varepsilon_{mned} = -\delta^{abc}_{mne} = \sum_{\sigma} {\rm sign}(\sigma) \,\delta_m^{\sigma_a}\,\delta_n^{\sigma_b}\,\delta_e^{\sigma_c} \,,\qquad \sigma\mbox{ runs over the permutation group }S_3 $$

The particular case $\,u^a= (0,0,0,1)\,$ corresponds to the laboratory frame and the relations (\ref{e10}) and (\ref{e13}) yield $E_a=(E_1,E_2,E_3,0)$ and $B^a=(B^1,B^2,B^3,0)$. This is why we respectively call $\,E_a\,$ and $\,B^d\,$ the electric field and the magnetic induction in the reference frame characterized by $u^a$.

We can proceed similarly with the skewsymmetric displacement tensor $H^{ab}$ and have
\begin{eqnarray}   \label{e14}
 & & H^{ab} = 2\,u^{[a} D^{b]} + \hat{H}^{ab}  \,, \qquad \qquad    \hat{H}^{ab}= \varepsilon^{abcd} u_c H_d \,, \qquad \qquad  u_a D^a = u^d H_d = 0 \\[2ex] \label{e14a}
 & & D^a = H^{ab} u_b  \qquad {\rm and} \qquad H_d = \frac12\, \varepsilon_{cdab} u^c H^{ab}
\end{eqnarray}
where $\,D^a\,$ and $H_d\,$ respectively stand for the electric displacement and the magnetic field in the reference frame characterized by the 4-velocity $u^a\,$.

As the constitutive relations only hold in the proper reference frame, we have that 
$$ D^a = \varepsilon \,E^a \qquad \quad{\rm and} \qquad \quad H_d = \frac1\mu\, B_d $$ 
provided that $u^a$ is the 4-velocity of the medium. Now, using (\ref{e13a}) and (\ref{e14a}), this amounts to
$$ H^{ab} u_b = \varepsilon \,F^{ab} u_b  \qquad \quad{\rm and} \qquad \quad
\varepsilon_{cdab} u^c H^{ab} = \frac1\mu\,\varepsilon_{cdab} u^c F^{ab}  $$
whence, after a little algebra, it follows that for an isotropic medium
\begin{equation}   \label{e15}
  H^{ab} = M^{abcd}F_{cd} \qquad {\rm with} \qquad M^{abcd}=\mu^{-1}\,\hat\eta^{a[c}\hat\eta^{d]b} + 2 \varepsilon\, u^{[a}\hat\eta^{b][c} u^{d]} 
\end{equation}
where $\,\hat\eta^{ab} = \eta^{ab}+ u^a u^b\,$ is the projector onto the hyperplane orthogonal to $u^b$.
The coefficients $M^{abcd}$ present the obvious symmetries
\begin{equation}   \label{e15a}
 M^{abcd} = - M^{bacd} =  - M^{abdc} =  M^{cdab}  
\end{equation}

\subsection{Minkowski energy-momentum tensor \label{S2.1}}
By a simple algebraic manipulation of Maxwell equations (\ref{e12a})
---which is quite similar to the proof of  Poynting theorem--- Minkowski obtained a local conservation law, namely
\begin{eqnarray} 
\partial_b \left(F^{ac} H^b_{\;c}\right) & = & - F^{ac} J_c + \partial_b F^a_{\;c} H^{bc} = - F^{ac} J_c + \frac12\, H^{bc} \left( \partial_b F^a_{\;c} +  \partial_c F_b^{\;a} \right) \nonumber  \\[2ex]   \label{e16a}
   & = & - F^{ac} J_c + \frac12\, H^{bc}  \partial^a F_{bc} = - F^{ac} J_c + \frac14\, \partial^a \left(H^{bc} F_{bc} \right)
\end{eqnarray}
where we have used the identity 
$$   H^{bc}  \partial^a F_{bc} = M^{bcmn} F_{mn} \partial^a F_{bc} = \frac12\,\partial^a\left(M^{bcmn} F_{mn} F_{bc} \right) = \frac12\,\partial^a \left(H^{bc} F_{bc} \right) $$
and the fact that $ M^{bcmn}$ is constant and presents the symmetries (\ref{e15a}).

Therefore the tensor 
\begin{equation} \label{e16}
  \Theta^{ab} = F^{ac} H^b_{\,\;c} - \frac14 \,\eta^{ab}\,H^{mn} F_{mn}
\end{equation}
fulfills the equation
\begin{equation} \label{e16b}
  \partial_b \Theta^{ab} = - F^{ac} J_c
\end{equation}
If there are no free charges, this relation becomes a local conservation law for the tensor $\Theta^{ab}$, which is called {\em Minkowski energy-momentum tensor}. It is generally non-symmetric, except if $\,\varepsilon\mu = 1\,$, in which case the tensor coefficient $M^{abcd}$ in (\ref{e15}) does not depend on $u^a\,$. This non-symmetry is at the origin of the so called Abraham-Minkowski controversy \cite{AbrahamMinkowski}

Contrarily, if there are free charges, the Lorentz force on the charges contained in the elementary volume  $\,\D^4 x\,$ 
is the result of the energy-momentum current flowing into it through its boundary
$$\, F^{ac} J_c  = - \partial_b \Theta^{ab} \,.  $$

\subsection{Derivation of Minkowski energy-momentum tensor from Noether theorem   \label{S2.2}}
The finding of the Minkowski energy-momentum tensor for non-dispersive media involves a certain amount of good luck. We shall now present an alternative derivation which is based in a Lagrangian formulation of Minkowski electrodynamics \cite{Ramos2015},\cite{AAA} and Noether theorem \cite{NoetherText}. The methodology is rather routine and involves very little creativity, which makes it appropriate for a further extension to the general case of dispersive media, as the one we shall endeavour in Section 3.

The configuration space variables are the 4-potencial components $A_b$ and the Lagrangian density in the absence of free charges is
\begin{equation} \label{e17}
 \mathcal{L} = \frac14\,M^{abcd}F_{ab} F_{cd}                                     
\,,\qquad \qquad \qquad  F_{ab} =   A_{b;a} - A_{a;b}
\end{equation}
where a `semi-colon' $\, ;\,$ means ``partial derivative'' and $\,M^{abcd}\,$ is constant for non-dispersive media and is given by (\ref{e15}). 

We shall need
$$ \frac{\partial\mathcal{L}}{\partial A_{a;b} } =  M^{abmn}\,F_{nm} = - H^{ab} \,, \qquad \qquad 
\frac{\partial\mathcal{L}}{\partial A_a } = 0     $$              
and the field equations $\,\displaystyle{ \frac{\partial\mathcal{L}}{\partial A_a } - \partial_b\left(\frac{\partial\mathcal{L}}{\partial A_{a;b} }\right) = 0 }\,$ are
\begin{equation} \label{e17z}
 \partial_b H^{ab} = 0 
\end{equation}

As the Lagrangian (\ref{e17}) is invariant under spacetime translations, by Noether theorem ---see the Appendix for an outline--- it has associated four conserved currents which conform the canonical energy-momentum tensor that, according to equation (\ref{N7z}), is
\begin{equation}   \label{e17a}
\mathcal{T}^{\;a}_c =  H^{ab} A_{b;c} - \frac14\, H^{mn} F_{mn} \delta^a_c  \,, \qquad \qquad  \partial_a \mathcal{T}^{\;a}_c = 0
\end{equation}
Besides of being non-symmetric (when both indices are raised), it is gauge dependent due to the occurrence of $\,A_{b;c}\,$ in the first term.

The angular momentum current that Noether theorem associates to infinitesimal Lorentz transformations ---see equations (\ref{N6}) and (\ref{N7z}) in the Appendix--- is
\begin{equation}   \label{e17b}
 \mathcal{J}^{\;\;\;b}_{ca} = 2 x_{[c} \mathcal{T}^{\;b}_{a]} + \mathcal{S}^{\;\;\;b}_{ca}  \quad\qquad {\rm where} \qquad \quad \mathcal{S}^{\;\;\;b}_{ca} = - 2\,A_{[c}H_{a]}^{\;\,b}   
\end{equation}
is the spin current. However, as $\,\mathcal{L}\,$ is not Lorentz invariant ---because it contains the 4-vector $\,u^a\,$ through the dielectric tensor $\,M^{abcd}\,$---, the angular momentum current is not locally conserved, $\,  \partial_b \mathcal{J}^{\;\;\;b}_{ca} \neq 0 $\,.

Then applying the symmetrization technique \cite{Dixon}, \cite{Landau} described in the Appendix  ---equation (\ref{N9})--- we obtain the so called Belinfante-Rosenfeld energy-momentum tensor 
\begin{equation}  \label{A17c}
\Theta^{ca} = \mathcal{T}^{ca} + \partial_b \mathcal{W}^{bac} \,, \qquad   {\rm where} \qquad  \mathcal{W}^{cab}  = \frac12\,\left(\mathcal{S}^{cab} +\mathcal{S}^{cba} -\mathcal{S}^{abc}  \right) \,, 
\end{equation}
that is $\, \mathcal{W}^{cab} = H^{ca} A^b \,$ and 
\begin{equation}  \label{e17d}
\Theta^{ca} = H^{ab} F^c_{\,\;b} - \frac14 \,\eta^{ca}\,H^{mn} F_{mn}  \,,  
\end{equation}
which recovers the  Minkowski energy-momentum tensor (\ref{e16}), \cite{AAA}, \cite{Ramos2015}. 
The fact that $\,W^{bac} = -W^{abc}\,$ implies that the local conservation  $\,\partial_a\Theta^{ba}= 0\,$ is a straight consequence of $\,\partial_a\mathcal{T}^{ba}= 0\,$.

Of course it is not symmetric, and it does not have to. Recall that the Belinfante tensor $\,\Theta^{ca}\,$ is symmetric if the angular momentum current $\,\mathcal{J}^{\;\;\;b}_{ca}\,$ is conserved \cite{Dixon}, which would follow from Noether theorem and the Lorentz invariance of the Lagrangian. Now the Lagrangian (\ref{e17}) is not Lorentz invariant, as commented above, because the dielectric tensor $\,M^{abcd}\,$ privileges the time vector $\,u^a\,$, which breaks boost invariance. As a matter of fact the Lagrangian (\ref{e17}) is invariant under the Lorentz subgroup that preserves, $\,u^a\,$, and it can be easily checked that the part of $\Theta^{ca} $ that is orthogonal to $u^b$ is indeed symmetric.

\section{Minkowski electrodynamics for dispersive media  \label{S3}}
When dealing with homogeneous isotropic dispersive media, the simple constitutive relations for constant $\varepsilon$ and $\mu$ must be replaced with the convolutions (\ref{e5}) or, in tensor spacetime form, the relations (\ref{e15}) are superseeded by 
\begin{equation}   \label{e18}
  H^{ab} = \tilde{M}^{abcd} \ast F_{cd} \qquad {\rm with} \qquad \tilde{M}^{abcd} = (2\pi)^{-2}\,\left[\tilde{m}(x) \,\hat\eta^{a[c}\hat\eta^{d]b} + 2 \tilde{\varepsilon}(x) 	\, u^{[a}\hat\eta^{b][c} u^{d]} \right]  
\end{equation}
where $\, \tilde{m}$ and $\tilde\varepsilon$ are the Fourier transforms of $\mu^{-1}$ and $\varepsilon\,$. 

Deriving the conservation equations for some energy-momentum current of the field in the Minkowski way, as in Section \ref{S2.1}, involves a trial and error game with an uncertain outcome. Alternatively we shall go for an extension of the method applied in Section \ref{S2.2}, namely (a) proposing a Lagrangian density from which the field equations are derived, then (b) obtaining the canonical energy-momentum and angular momentum currents by application of Noether theorem and finally (c) applying the symetrization technique \cite{Belinfante}, \cite{Dixon}, \cite{Landau} to derive a Belinfante-Rosenfeld energy-momentum tensor. 

The constitutive relations (\ref{e18}) are non-local and so are the field equations, therefore we postulate the non-local action integral
\begin{equation}   \label{e19}
 S = \int \D^4 x\,\int \D^4 y\, \frac1{4}\,\tilde{M}^{abcd}(x- y)\,F_{ab}(x) \,F_{cd}(y)  
\end{equation}
where 
$F_{ab} = A_{b;a} - A_{a;b} \,$, 
$\tilde{M}^{abcd}(x)$ is skewsymmetric in both pairs of indices $\tilde{M}^{abcd}(x)  = - \tilde{M}^{bacd}(x)  = - \tilde{M}^{abdc}(x) $ and
\begin{equation}   \label{e19a}
\tilde{M}^{abcd}(-x) = \tilde{M}^{cdab}(x) 
\end{equation}
which, particularized to the specific form (\ref{e18}), amounts to require that $\tilde{m}$ and $ \tilde\varepsilon\,$ are even functions,
\begin{equation}   \label{e19aa}
\tilde{m}(-x) = \tilde{m}(x) \qquad\quad {\rm and} \qquad\quad \tilde\varepsilon(-x) =\tilde\varepsilon(x) \,,
\end{equation}
(Notice that if $\tilde{M}^{abcd}(x)$ contained an ``antisymmetric'' component, namely $\tilde{N}^{abcd}(x)=-\tilde{N}^{cdab}(-x)$, it would not contribute to the action integral (\ref{e19}).)
The action (\ref{e19}) also includes the non-dispersive case, i. e. $\tilde{M}^{abcd}(x-y)=M^{abcd}\,\delta^4(x-y)$, where $M^{abcd}$ is a constant tensor.

The Lagrangian density is $\displaystyle{ \mathcal{L} =  \frac1{4}\, F_{ab}(x) \,\int \D^4 y\, \tilde{M}^{abed}(y)\,F_{ed}(x-y) }$, or
\begin{equation}   \label{e20}
 \mathcal{L} = \frac1{4}\,F_{ab} \,H^{ab} \,, \qquad {\rm with} \qquad H^{ab} = \left(\tilde{M}^{abed}\ast F_{ed} \right) \,,
\end{equation}  
which is obviously non-local because $ \mathcal{L}(x)$ depends on the field derivatives $A_{b;a}(x)$ and, due to the convolution, it also depends on the  values $ A_{e;d}(y)$ at any other point.

As the derivation of Euler equations from a non-local Lagrangian is not a subject that one commonly founds in standard textbooks, we present the whole procedure in the Appendix. We also proof there a generalisation of Noether theorem that we shall use to obtain the energy-momentum tensor as the current associated to the invariance of the Lagrangian (\ref{e20}) under spacetime translations. The developments in the Appendix are fundamental to understand the guesswork in what follows, but the reader can skip them.

In order to not distracting the reader from the main thread of the paper, we start from the Maxwell equations (\ref{A11}) for the dispersive medium
\begin{equation}  \label{e21}
\partial_a H^{ab} = 0 
\end{equation}
and guess the two energy-momentum tensors (this is not a blind guess but oriented by the full work in the Appendix) ---see equations (\ref{N5}) and (\ref{D4A})---\\[1ex]
{\bf the canonic energy-momentum current} 
\begin{equation}   \label{D1}  
\mathcal{T}^{ab} = -\mathcal{L}\,\eta^{ab} + \frac12\,H^{bc}\,A_c^{\; ;a} - \frac12\,\int_{\mathbb{R}^4} \D\xi\,\xi^b \int_0^1\D\lambda\,
F_{fe}(X-\xi)\,\tilde{M}^{dnfe}_{\quad\; ;d}(\xi)\,A_n^{\; ;a} (X) 
\end{equation}
and 
{\bf the Belinfante-Rosenfeld tensor}
\begin{eqnarray}
\Theta^{ba}  &=& \frac12\, H^{ca} F_c^{\;\,b} -\frac14\,\eta^{ab} F_{ed} H^{ed} + 
\frac12\,F_{fe} \left[\tilde{M}^{fed[b}\ast F^{a]}_{\;\,d} + 
\tilde{M}^{fed(b;a)} \ast A_d + \frac12 \left( y^b \tilde{M}^{fedn}\right)\ast F_{dn}^{\;\;;a} \right] \nonumber \\[2ex]
 & & - \frac12\, \int \D^4 y\tilde{M}^{ndfe}(\xi)\,\frac{\partial\;\,}{\partial \xi^d}\,\int_0^1 \D\lambda \,\xi^{(a}\left\{ F_{fe}(X-\xi) \left[A_n^{\;;b)}(X) + F^{b)}_{\;\,n}(X)\right] \right. \nonumber\\[2ex] \label{D4}
 & &  - F_{fe;n}(X- \xi) A^{b)}(X)  + \left. \delta^{b)}_n \left[F_{fe}(X- y) A^c(X) \right]_{;c} - \lambda  \xi^{b)} \left[F_{fe}(X- \xi) A_n^{\;;c}(X)  \right]_{;c} \right\}
\end{eqnarray}
where $X = x + \lambda \xi\,$.

Since both expressions depend linearly on the electromagnetic potential, they are gauge dependent. 
The potential $A_b$ can be eliminated by means of the inverse of the definition $F_{ed} = A_{d;e} - A_{e;d}\,$; indeed, by the Poincar\'e Lemma \cite{Spivak} we have that
\begin{equation}   \label{D1a}  
 A_b(x) = \int_0^1 \D\tau\, \tau\,x^c\, F_{cb}(\tau x) + \partial_b f(x) 
\end{equation}
where $\,f(x)\,$ is an arbitrary function that is related with gauge transformations. Thus, due to the linear dependence, both energy-momentum tensors split in one part that only depends on $F_{cd}$ and is gauge independent, and another one that depends linearly on $\,\partial_e f$, i. e. a gauge dependent contribution. 
It can be easily proved that the gauge parts are conserved, hence we can take the gauge independent parts as the definitions of the energy-momentum tensors.  

Let us now check whether these tensors are locally conserved. To begin with we have that, by its construction,
$$ \partial_b\Theta^{ab} = \partial_b\mathcal{T}^{ab} \,.$$
As for the conservation of the canonic tensor, from (\ref{D1}) we have that
$$ \partial_b\mathcal{T}^{ab} = - \frac14 F_{cd} H^{cd;a} - \frac12\,\int_{\mathbb{R}^4} \D\xi\,\tilde{M}^{dnfe}_{\quad\; ;d}(\xi)\,
\int_0^1\D\lambda\, \xi^b \,\partial_b\left[F_{fe}(x+[\lambda-1]\xi) A_n^{\; ;a} (x+\lambda\xi) \right] \,,$$
where we have used the field equations (\ref{e21}). 

Including now the identity
$$ \xi^b \,\partial_b\left[F_{fe}(x+[\lambda-1]\xi) A_n^{\; ;a} (x+\lambda\xi) \right] = \frac{\partial\;}{\partial\lambda}\,\left[F_{fe}(x+[\lambda-1]\xi) A_n^{\; ;a} (x+\lambda\xi) \right] \,,$$
we can perform the integral and arrive at
\begin{eqnarray*}
 \partial_b\mathcal{T}^{ab} &=& - \frac14 F_{cd} H^{cd;a} - \frac12\,\int_{\mathbb{R}^4} \D\xi\,\tilde{M}^{dnfe}_{\quad\; ;d}(\xi)\,
\left[F_{fe}(x) A_n^{\; ;a} (x+\xi) - F_{fe}(x-\xi) A_n^{\; ;a} (x)\right] \\[2ex]
 &=& - \frac14 F_{cd} H^{cd;a} - \frac12\,\int_{\mathbb{R}^4}\D\xi\,\tilde{M}^{dnfe}_{\quad\; ;d}(\xi)\,F_{fe}(x) A_n^{\; ;a}(x+\xi) + \frac12\,H^{dn}_{\;\;\, ;d} A_n^{\;\,;a}
\end{eqnarray*}
The last term vanishes due to the field equations and, integrating by parts and using the $dn$-skewsymmetry, we have that
$$ \partial_b\mathcal{T}^{ab} = - \frac14 F_{cd} H^{cd;a} + \frac14\,F_{fe} \,\int_{\mathbb{R}^4}\D \xi\,\tilde{M}^{dnfe}(-\xi)\,F_{dn}^{\;\; ;a}(x-\xi) \,,$$
that is
\begin{equation}  \label{D5}
 \partial_b\mathcal{T}^{ab}  = -\frac12\,F_{fe} \,\tilde{M}_-^{fedn} \ast F_{dn}^{\quad ;a} \qquad {\rm where} \qquad 
\tilde{M}_-^{cdef}(\xi) = \frac12\,\left[\tilde{M}^{cdef}(\xi) - \tilde{M}^{efcd}(-\xi) \right]
\end{equation}
In case that the field equations (\ref{e21}) can be derived from a Lagrangian, then the symmetry relation (\ref{e19a}) implies that $\tilde{M}_-^{fedn}$ vanishes and both energy-momentum tensors, (\ref{D1}) and (\ref{D4}), are locally conserved. 

\subsection{Real dispersive media: absorption and causality  \label{S3.1}}
Recall that the symmetry (\ref{e19a}) implies that the dielectric and magnetic functions are even functions. Hence their Fourier transforms $\varepsilon$ and $\mu$ are real valued and so is the refractive index $\,n\,$ as well.
However causality implies that the real and imaginary parts of the functions $\varepsilon$ and $\mu$ must fulfill either the Kramers-Kr\"onig relations \cite{KramersKronig} (in the optical approximation, i. e. $\varepsilon$ and $\mu$ only depend on the angular frequency $\omega$) or the Leontovich relations \cite{Leontovich}, \cite{LlosaSalvat} in the general case. 
As a consequence, were $\varepsilon$ and $\mu$ real valuated, they should be constant and the medium would be non-dispersive.

For a real dispersive medium $\varepsilon$ and $\mu$ are not constant, therefore the symmetry relation (\ref{e19a}) is not fulfilled and the right hand side of (\ref{D5}) does not vanish. However, including (\ref{e18}), we have
$$\tilde{M}_-^{fedn} = (2\pi)^{-2}\,\left[\tilde{m}_- \,\hat\eta^{f[d}\hat\eta^{n]e} + 2 \tilde{\varepsilon}_-\, u^{[f}\hat\eta^{e][d} u^{n]} \right] \,,$$
with $\quad \displaystyle{ \tilde{m}_-(y) = \frac{\tilde{m}(y) - \tilde{m}(-y)}2 \quad}$ and similarly for $\, \tilde\varepsilon_-(y)\,$, and using (\ref{e13}) to separate the electric and magnetic parts, we can write (\ref{D5}) as
$$ \partial_b\mathcal{T}^{ab}  = (2\pi)^{-2}\,\left[ E_d\,\left(\tilde\varepsilon_-\ast E^{d;a}\right) - B_d\,\left(\tilde{m}_-\ast B^{d;a}\right) \right] $$ 
Now, as the Fourier transforms of $\,\tilde{m}_-(x)\,$ and $\, \tilde\varepsilon_-(x)\,$ are connected to $\imap \,m(k)$ and $\imap\, \varepsilon(k)$, i.e. the absorptive parts of the magnetic and dielectric functions, we have that the failure of local conservation of energy-momentum in a real medium is due to absorption.

\subsection{Plane wave solutions \label{S3.4}}
These are particular solutions of the Maxwell equations in the form
$$F_{cd} = f_{cd} e^{i k_b x^b}\,,\qquad {\rm with} \qquad f_{cd} +f_{dc} = 0\,$$ 
that, substituted in the field equation (\ref{e21}), yields
\begin{equation}   \label{OP1}  
M^{abcd} (k)\, f_{cd} k_b = 0 
\end{equation}
where $M^{abcd} (k)$ is the Fourier transform of $\tilde{M}^{abcd} (k)$.
Now as $F_{cd}$ can be derived from an electromagnetic potential $A_b$, it must fulfill the first pair of Maxwell equations which for plane waves reads $\, k_b f_{cd} +  k_c f_{db} +  k_d f_{bc} = 0 \,$ and whose general solution is   
\begin{equation}   \label{OP2}  
 f_{cd} = f_c k_ d - f_d k_c
\end{equation}
where $f_c$ is the wave polarization vector and is determined apart from the addition of a multiple of $k_c\,$.
Substituting this into equation (\ref{OP1}), we arrive at
\begin{equation}   \label{OP3}  
M^{abcd}(k) \, k_b \, k_d \, f_c = 0 
\end{equation}
which is a linear homogeneous system and admits non-trivial solutions for the polarization vector if, and only if, 
$$   \det\left[ M^{abcd}(k) \, k_b \, k_d \right] = 0\,.  $$

We shall assume that the dielectric tensor $\tilde{M}^{abcd}$ has the form (\ref{e18}), hence its Fourier transform is
\begin{equation}   \label{OP6}  
 M^{abcd}(k) = m(k) \,\hat\eta^{a[c}\hat\eta^{d]b} + 2  \varepsilon(k)\, u^{[a}\hat\eta^{b][c} u^{d]} 
\end{equation}
If the medium is spatially isotropic, the functions $\varepsilon(k) $ and $\,m(k)=\mu^{-1}(k) \,$ depend on the wave vector $k^b$ through the scalars $\omega = - k^b u_b$ and $q^2:= k^b k_b + \omega^2\,$, where we have taken  
$$ k^a = \omega u^a + q\,\hat{\mathbf{q}}^a \,,   \qquad {\rm with} \qquad \hat{\mathbf{q}}^a\hat{\mathbf{q}}_a =  1 \,, \qquad \quad \hat{\mathbf{q}}^a u_ a = 0 $$
As the polarization is determined up to the addition of a multiple of $k^c$, we can choose it so that
$$ f^c = \psi\,\hat{\mathbf{q}}^c + f_\perp^c \qquad {\rm with} \qquad f_\perp^c u^c = f_\perp^c \hat{\mathbf{q}}^c = 0 \,.$$
Substituting this in equation (\ref{OP3}), we arrive at
$$ \left( q^2 - \omega^2  n^2  \right)\,f_a^\perp  = 0 \,, \qquad \quad \psi=0 \,,$$
where $\,n(q,\omega) =\sqrt{\varepsilon \mu}\,$ is the refractive index.

Hence there are non-trivial solutions if, and only if,
\begin{equation}     \label{OP4}
q^2 - \omega^2  n^2   = 0 \qquad \qquad {\rm and}  \qquad \qquad  f_b k^b = f_b u^b  = 0   
\end{equation}

Maxwell equations thus imply that: 
\begin{description}
\item[(a)] the waves are polarized transversely to the plane spanned by $k^b$ and $u^b$ and 
\item[(b)] the phase velocity satisfies the dispersion relation $\,q = \omega\,n(q,\omega) \,$.
\end{description}

\subsection{The general solution of field equations \label{S3.2}}
As it is well known \cite{Barnaby}, \cite{Calcagni}, the initial data problem for a non-local PDS like the Maxwell equations (\ref{e21}) is not as simple as the Cauchy problem for a partial differential system of first order, as the field equations for non-dispersive media. We shall now see that the nature of the initial data problem for (\ref{e21}) depends on whether the number of real roots of the dispersion relations (\ref{OP4}) is finite or not.

As Maxwell equations (\ref{e12a}) and (\ref{e21}) are linear and they involve a convolution, the Fourier transform is a good tool to solve them. We write 
\begin{equation}     \label{GS1}
F_{cd}(x) = (2\pi)^{-2} \,\int \D k \,f_{cd}(k)\,e^{ik_a x^a} 
\end{equation}
that, substituted in Maxwell equations, yields
\begin{equation}     \label{GS2}
f_{cd} = f_c k_d - f_d k_c \qquad {\rm and} \qquad M^{abcd}(k) \, k_b \, k_d \, f_c = 0 
\end{equation}
which, as discussed above means that $f_c$ vanishes unless the dispersion relation (\ref{OP4}) is fulfilled, i. e.
$$ f_{cd}(k) = f^+_{cd}(k)\,\delta\left[q - \omega \,n(q,\omega)\right] + f^-_{cd}(k)\,\delta\left[q + \omega \,n(q,\omega)\right]  $$
or
\begin{equation}     \label{GS3}
 f_{cd}(k) = \sum_\alpha f^{\alpha}_{cd}(\mathbf{q})\,\delta\left[\omega-\omega_\alpha(\mathbf{q})\right] + \sum_\beta f^{\beta}_{cd}(\mathbf{q})\,\delta\left[\omega-\omega_\beta(\mathbf{q})\right] \, , 
\end{equation}
where $\omega_\alpha$ (resp. $\omega_\beta$) are the positive (resp. negative) real roots of the dispersion relation (\ref{OP4}).

The arbitrary coefficients $ f^{\alpha}_{cd}(\mathbf{q}) $ and $ f^{\beta}_{cd}(\mathbf{q})$ depend on the initial data. Indeed, the $n$-th time derivative of (\ref{GS1}) at $x^a= (0,\mathbf{x})$ yields
$$ \partial^n_t F_{cd}(0,\mathbf{x}) = (2\pi)^{-2} \,\int \D \mathbf{q} \,e^{i\mathbf{q}\cdot\mathbf{x}} \,\left[\sum_\alpha f^{\alpha}_{cd}(\mathbf{q})\,(-i\omega_\alpha)^n + \sum_\beta f^{\beta}_{cd}(\mathbf{q}) \,(-i\omega_\beta)^n \right]\,, $$
where (\ref{GS3}) has been used to perform the integration on $\omega$, whence it follows that
\begin{equation}     \label{GS4}
\sum_\alpha f^{\alpha}_{cd}(\mathbf{q})\,\omega^n_\alpha + \sum_\beta f^{\beta}_{cd}(\mathbf{q}) \,\omega^n_\beta = \frac{i^n}{2\pi}\, 
\int \D \mathbf{x} \,e^{-i\mathbf{q}\cdot\mathbf{x}} \, \partial^n_t F_{cd}(0,\mathbf{x}) 
\end{equation}
Since we have as many unknowns $f^{\alpha}_{cd}$ and $f^{\beta}_{cd}$ as the number of real roots of the dispersion relation (\ref{OP4}), this is the number of initial ($t=0$) time derivatives of $F_{cd}(t,\mathbf{x}) $ that are needed at least to determine a solution. 
If the number of real roots is $N <\infty\,$, then the giving of $\,A_d(\mathbf{x},0) \,$ and its time derivatives up to the order $N-1$ determines the solution of the system; otherwise the problem of initial data requires further study.

\subsection{The energy-momentum tensor for a wave packet \label{SS4.1}}
Aiming to compare the energy-momentum tensor obtained here with other proposals advanced in the literature, e. g.  \cite{Landau2}, \cite{Jackson2} and \cite{Schwinger}, we shall particularize the expression (\ref{D4}) of the Belinfante tensor to the wave packet 
\begin{equation}     \label{OP5}
F_{cd}(x) =  \realp \left(\underline{F}_{cd}(x) \, e^{ik_a x^a} \right) = |\underline{F}_{cd}(x)|\,\cos\left(k_a x^a + \varphi_{cd}\right)    \,, \qquad \qquad k^a = \mathbf{q}^a + \omega u^a
\end{equation}
where $\underline{F}_{cd}(x)$ is a ``slowly'' varying complex amplitude (if compared with the rapidly oscillating carrier $\,e^{ik_c x^c}\,$), $|\underline{F}_{cd}(x)|$ is the modulus of each component and $\varphi_{cd}$ is the phase.
Furthermore, we shall restrict to the optical approximation, that is $\varepsilon$ and $\mu$ only depend on the frequence $\,\omega = - u_a k^a\,$, therefore we shall take
\begin{equation}     \label{OP7a}
 \tilde{M}^{abcd}(y) = (2\pi)^{-1/2}  \delta^3(\mathbf{y})\, \tilde{m}^{abcd}(\tau) \,,\qquad {\rm with} \qquad 
\tilde{m}^{abcd} = \tilde{m}(\tau) \,\hat\eta^{a[c}\hat\eta^{d]b} + 2 \tilde{\varepsilon}(\tau) 	\, u^{[a}\hat\eta^{b][c} u^{d]} \,,
\end{equation} 
$\tau= -y^a u_a\,$ and $\,\mathbf{y}^a = y^a - \tau u^a\,$ is the spatial part of $y^a\,$.

Using this, the displacement tensor (\ref{e18}) yields
\begin{equation}     \label{OP7}
 H^{ca}(x) =  \realp \left(\underline{H}^{ca}(x) \, e^{ik_b x^b } \right)
  \qquad {\rm with} \qquad  \underline{H}^{ca}(x) \approx \tilde{m}^{caed}(\tau) \underline{F}_{ed}(x)
\end{equation}
where $\,\approx\,$ means that the ``slow variation'' approximation has been included to evaluate the convolution, that is
$$ (2\pi)^{-1/2}  \int \D\tau\, \tilde{m}^{caed}(\tau) \underline{F}_{ed}(x^b-\tau u^b)\,e^{ik_b x^b + i \omega \tau } \approx \tilde{m}^{caed}(\tau) \underline{F}_{ed}(x)\,e^{ik_b x^b} \,.$$

Using (\ref{OP7}), the Maxwell equations become
$$ D_c \underline{H}^{ca} \approx 0 \qquad {\rm and} \qquad D_b\underline{F}_{cd}+ D_c\underline{F}_{db}+ D_d\underline{F}_{bc} = 0 $$
where $\,D_b = \partial_b + i k_b\,$.
For slowly varying amplitudes they reduce to the Maxwell equations for a plane wave and we can write (see Section \ref{S3.4})
\begin{equation}     \label{OP8}
\underline{F}_{cd} \approx -\frac1\omega\,\left( \underline{E}_c k_d - \underline{E}_d k_c \right) \qquad{\rm and}\qquad m^{caed} k_c \underline{E}_e k_d \approx 0
\end{equation}
where $\,\underline{E}_c = \underline{F}_{cd} u^d\,$ is the electric field and $m^{caed}(\omega)$ is the Fourier transform of $\tilde{m}^{caed}(\tau)\,$. Similarly as in Section \ref{S3.4}, the second equation implies that
\begin{equation}     \label{OP8a}
\underline{E}_c k^c = 0   \qquad \qquad{\rm and}\qquad \qquad q^2 = \omega^2 \varepsilon(\omega) \mu(\omega)
\end{equation}
Moreover, from (\ref{OP5}) and (\ref{OP8}) it follows that the electromagnetic potential is 
\begin{equation}     \label{OP9}
 A_b(x) =  \realp \left(\underline{A}_b(x) \, e^{ik_cx^c} \right)
  \qquad {\rm with} \qquad  \underline{A}_c \approx - \frac{i}\omega \,\underline{E}_c + \underline\alpha\,k_c \,,
\end{equation}
where $\,\underline\alpha(x)\,$ is an arbitrary gauge function.

If we now substitute the wave packet (\ref{OP5}) in the Belinfante tensor expression (\ref{D4}), we find that:
\begin{itemize}
\item The evaluation of the convolution products in the first line yields
$$ \tilde{M}^{edh[b}\ast F^{a]}_{\;\,h} \approx \realp\left( m^{edh[b}(\omega) \underline{F}^{a]}_{\;\,h} e^{i k_c x^c}   \right) \,, \qquad \quad
\tilde{M}^{edh(b;a)}\ast A_h \approx \realp\left( m^{edh(b}(\omega) D^{a)}\underline{A}_h e^{i k_c x^c}   \right) $$
$$ {\rm and} \hspace*{8em}  \left( y^b \tilde{M}^{edhf}\right)\ast F_{hf}^{\;\;;a} \approx \realp\left(-i\,D^a \underline{F}_{hf} m^{\prime\,edhf}(\omega)    \right) \hspace*{8em}$$
where a ``prime'' means derivative with respect to $\omega$. Thus every term in the first line of (\ref{D4}) has the form
\begin{equation}  \label{OP10}
\Phi\,\Psi = \realp\left(\underline{\Phi} e^{i k_c x^c}\right)\,\realp\left( \underline{\Psi} e^{i k_c x^c}   \right) = \frac12\, \realp\left(\underline{\Phi}\,\underline{\Psi} e^{2 i k_c x^c} + \underline{\Phi}\,\underline{\Psi}^\ast \right) 
\end{equation}
and consists of a slowly varying part plus a rapidly oscillating one. Taking the average over a period of the  carrier we obtain
$$ \langle \Phi\,\Psi \rangle = \frac12\, \realp\left(\underline{\Phi}\,\underline{\Psi}^\ast \right) \,.$$
This is the physically meaningful quantity as far as the carrier period lasts much less than a physical measurement.

\item To evaluate the integral in the second and third lines in (\ref{D4}) we realize that each term contains a group of the kind of (\ref{OP10}) and a Fourier integral of $\,\tilde{m}^{hfed}(\tau)\,$. We shall use the slow variation approximation and take the mean over a carrier period.
\end{itemize}

A tedious calculation leads to 
\begin{eqnarray*}
\langle \Theta^{ba} \rangle &\approx & \frac14\,\realp \left[2 \underline{H}^{ca} \underline{F}_c^{\ast \,b} - \underline{H}^{c[a} \underline{F}_c^{\ast \,b]} - \frac12\,\underline{H}^{cd} \underline{F}_{cd}^\ast \,\eta^{ab} + \underline{F}_{ed}^\ast  m^{edh[b} \underline{F}^{a]}_{\;\;h}   \right. \\[2ex]
  &  & + i\,m^{edh(a} k^{b)} \left( \underline{F}_{ed}^\ast \underline{A}_h - \underline{F}_{ed}  \underline{A}_h^\ast \right)- \frac12\,u^a k^b \underline{F}_{ed}\, m^{\prime \;edhf} \underline{F}_{hf}^\ast
\end{eqnarray*}
which, using the relations (\ref{OP8}-\ref{OP9}), can be simplified to
\begin{equation}  \label{OP11}
\langle \Theta^{ba} \rangle \approx \frac12\,\realp \left[\underline{H}^{ca} \underline{F}_c^{\ast \,b}  - \frac14\,\underline{H}^{cd} \underline{F}_{cd}^\ast \,\eta^{ab} - \frac14\,u^a k^b \underline{F}_{ed}\, m^{\prime \;edhf} \underline{F}_{hf}^\ast \right]
\end{equation}
From the latter we easily obtain that the energy density in the medium rest frame ($\, u^a =\delta_4^a\,$) is
$$ \mathcal{U}  = \langle \Theta^{44} \rangle \approx \frac14\,\realp \left[(\varepsilon + \omega\varepsilon^\prime) \underline{\mathbf{E}} \cdot \underline{\mathbf{E}}^\ast + (m - \omega m^\prime) \underline{\mathbf{B}} \cdot \underline{\mathbf{B}}^\ast \right] \,,  $$
where $\,\underline{\mathbf{E}} \cdot \underline{\mathbf{E}}^\ast = E_a E^{\ast\,a}\,$, or
\begin{equation}  \label{OP12}
\mathcal{U} \approx \frac14\,\realp \left[\frac{\D (\varepsilon \omega)}{\D\omega}\, \underline{\mathbf{E}} \cdot \underline{\mathbf{E}}^\ast + \frac{\mu^\ast}{\mu} \,\frac{\D (\mu \omega)}{\D\omega}\, \underline{\mathbf{H}} \cdot \underline{\mathbf{H}}^\ast \right]  
\end{equation}
which reproduces previous results in the literature ---see \cite{Landau2}, \cite{Jackson2} and \cite{Schwinger}.

The momentum density in the medium rest frame is $\,G^i = \langle \Theta^{i4} \rangle\,$ and, in an obvious vector notation, we obtain from (\ref{OP11}) that
$$ \mathbf{G} \approx \frac12\,\realp \left[\varepsilon\, \underline{\mathbf{E}} \times \underline{\mathbf{B}}^\ast + \frac12\,\left(\varepsilon^\prime \underline{\mathbf{E}} \cdot \underline{\mathbf{E}}^\ast - m^\prime \underline{\mathbf{B}} \cdot \underline{\mathbf{B}}^\ast \right)\, \mathbf{q} \right] $$
Now, from the Maxwell equations (\ref{OP8}) it follows that
$$ \underline{\mathbf{E}} \times \underline{\mathbf{B}}^\ast = \frac{\underline{\mathbf{E}} \cdot \underline{\mathbf{E}}^\ast}\omega\,\mathbf{q}
\qquad {\rm and} \qquad \underline{\mathbf{B}} \cdot \underline{\mathbf{B}}^\ast = \varepsilon\mu\,\underline{\mathbf{E}} \cdot \underline{\mathbf{E}}^\ast $$  
which substituted above yields
\begin{equation}  \label{OP13}
\mathbf{G} \approx \frac14\,\realp \left[\frac1{\omega\mu}\,\frac{\D (\varepsilon\mu \omega^2)}{\D\omega}\,\underline{\mathbf{E}} \times \underline{\mathbf{B}}^\ast \right]
\end{equation}

The Poynting vector is 
$$\,S^i = \langle \Theta^{4i} \rangle \approx \frac12\,\realp \left[\underline{\mathbf{E}}^\ast \times \underline{\mathbf{H}}^\ast \right]   $$
and the Maxwell stress tensor is
$$ T^{ij} = - \langle \Theta^{ij} \rangle \approx  \frac12\,\realp \left[\underline{E}^{\ast\,i}\underline{D}^j + \underline{H}^i \underline{B}^{\ast\,j} -\frac12\,\left(\underline{\mathbf{D}} \cdot \underline{\mathbf{E}}^\ast + \underline{\mathbf{H}} \cdot \underline{\mathbf{B}}^\ast   \right) \,\delta^{ij}\right] $$

\section{Conclusion}
We have tackled the derivation of an energy-momentum tensor for electromagnetic field in a linear, isotropic, homogeneous dispersive medium. Our set up is based on a quadratic Lagrangian for the electromagnetic field. Due to dispersivity this Lagrangian must be non-local, i. e. it depends on the field at several different points. In the non-dispersive limit, the Lagrangian becomes local and first order, and Minkowski theory \cite{Minkowski1908} is recovered. 

Homogeneity implies that the Lagrangian is invariant by spacetime translations. Hence the conservation of some energy-momentum current must follow from an eventual Noether theorem for non-local Lagrangians. As we are aware that this subject is not currently found in textbooks, we have devoted the Appendix to outline the derivation of both the field Euler-Lagrange equations and Noether theorem for a non-local Lagrangian.

As a result we have obtained an explicit expression for the canonical energy-momentum tensor $\,\mathcal{T}_b^{\;\, a}\,$ which depends quadratically and non-locally on the Faraday tensor and its first order derivatives. In the non-dispersive limit this tensor does not coincide with the Minkowski energy-momentum tensor; the difference is the 4-divergence of an antisymmetric tensor of order three. We have derived this correction by applying the Belinfante-Rosenfeld technique\cite{Dixon} and obtained an energy-momentum tensor $\,\Theta_b^{\;\, a}\,$ which in the non-dispersive limit does reduce to Minkowski tensor.    
In general the tensor $\,\Theta^{b a}\,$ is not symmetric, as Minkowski tensor is not either. This is due to the fact that the angular momentum current is not conserved because the Lagrangian is not Lorentz invariant, as expected because the rest reference system of the medium is a privileged one.

It must be said that our Lagrangian model has the disadvantage that its scope is restricted to non-absorptive media. Indeed, the action (\ref{e19}) implies the symmetry conditions (\ref{e19a}) and (\ref{e19aa}), whence it follows that $\varepsilon(\omega,k)\,$ is real for real $\omega $ and $k$, and it must be recalled that the absorptive behavior of a medium is connected with the imaginary part of its dielectric function $\varepsilon\,$. Moreover, if this imaginary part vanishes, it follows from Kramers-Kr\"onig relations that $\varepsilon$ and $\mu$ must be constant. Therefore, if the Lagrangian model does not violate causality, then it must be non-dispersive, i.e. local.

If we give up the Lagrangian model and base the description of the causal non-dispersive medium on Maxwell equations, we can stiil propose (\ref{D1}) and (\ref{D4}) as two possible definitions for energy-momentum currents, respectively the canonical and the Belinfante-Rosenfeld tensors. Evaluating then their 4-divergences provided that the field equations (\ref{e21}) hold, we than find that they are not locally conserved and this is due to the absorptive components of the dielectric and magnetic functions, i. e. $\imap(\varepsilon)$ and $\imap(\mu)$.

We have then specialized our Belinfante-Rosenfeld energy-momentum tensor to the electromagnetic field of slowly varying amplitude over a rapidly oscillating carrier wave, for a medium in the optical approximation ---that is $\varepsilon$ and $\mu$ only depend on the frequency $\omega$. Taking the average over one period of the carrier and using the slow motion approximation we have evaluated the energy and momentum densities, the Poynting vector and the Maxwell stress tensor in the rest reference frame. Energy density is the only of these quantities that are given in some textbooks adn our result agrees with them \cite{Landau2}, \cite{Jackson2}, \cite{Schwinger}.

\section*{Acknowledgment}
Funding for this work was partially provided by the Spanish MINCIU and ERDF (project ref. RTI2018-098117-B-C22).

\section*{Data availability}
The data that support the findings of this study are available within the article.
\renewcommand{\theequation}{A.\arabic{equation}}
\setcounter{equation}{0}
\section*{Appendix: Non-local Lagrangian field theories}
Consider the action integral
\begin{equation}   \label{A1}
 S = \int_{\mathbb{R}^4} \mathcal{L}\left([\phi^\alpha],x\right)\,\D x
\end{equation}
where the Lagrangian density $\mathcal{L}$ depends on all the values $\phi^\alpha(y)$ of the field variables at points $y$ other than $x$. This is why we refer to it as non-local and the Lagrangian density (\ref{e20}) is an example. 

The variation of the action is
$$ \delta S = \int_{\mathbb{R}^4} \D x\,\int_{\mathbb{R}^4} \D y\, \frac{\delta \mathcal{L}\left([\phi^\alpha],x\right)}{\delta \phi^\alpha(y)}\,\delta \phi^\alpha(y)$$
The requirement that this variation vanishes for any $\,\delta \phi^\alpha(y)$ with compact support leads to the Euler-Lagrange equation
\begin{equation}   \label{A2}
 \Pi_\alpha(y) \equiv \int_{\mathbb{R}^4} \D x\,\Lambda_\alpha(x,y) = 0 \,, \qquad {\rm where} \qquad \Lambda_\alpha(x,y) = 
\frac{\delta \mathcal{L}\left([\phi^\alpha],x\right)}{\delta \phi^\alpha(y)}
\end{equation}

\subsection*{Noether theorem}
Consider the infinitesimal transformation
\begin{equation}  \label{A3}
x^{\prime a} = x^a + \delta x^a \,, \qquad \quad \phi^{\prime \alpha}(x) = \phi^\alpha(x) + \delta \phi^\alpha(x) \,.
\end{equation}
Let $\mathcal{V}$ be a spacetime volume and  ${\mathcal{V}^\prime}$ its transformed according to (\ref{A3}) and define 
\begin{equation}  \label{A3a}
 \Delta S(\mathcal{V}) \equiv \int_{\mathcal{V}^\prime} \mathcal{L}\left([\phi^{\prime \alpha}],x^\prime\right)\,\D x^\prime -
\int_{\mathcal{V}} \mathcal{L}\left([\phi^\alpha],x\right)\,\D x 
\end{equation}
Replacing with $x$ the dummy variable $x^\prime$ in the first term on the right hand side we have that
\begin{equation}  \label{A4}
 \Delta S(\mathcal{V}) = \int_{\mathcal{V}^\prime} \mathcal{L}\left([\phi^{\prime \alpha}],x\right)\,\D x -\int_{\mathcal{V}} \mathcal{L}\left([\phi^\alpha],x\right)\,\D x  
\end{equation}

As depicted in Figure 1, the volumes $\mathcal{V}$ and $\mathcal{V}^\prime$ differ very little: they share a large common part  $\mathcal{V}_0$ and differ in an infinitesimal part near the boundary $\partial \mathcal{V}_0$
\begin{figure}  \label{f1}
\begin{center}
\includegraphics[width=12cm]{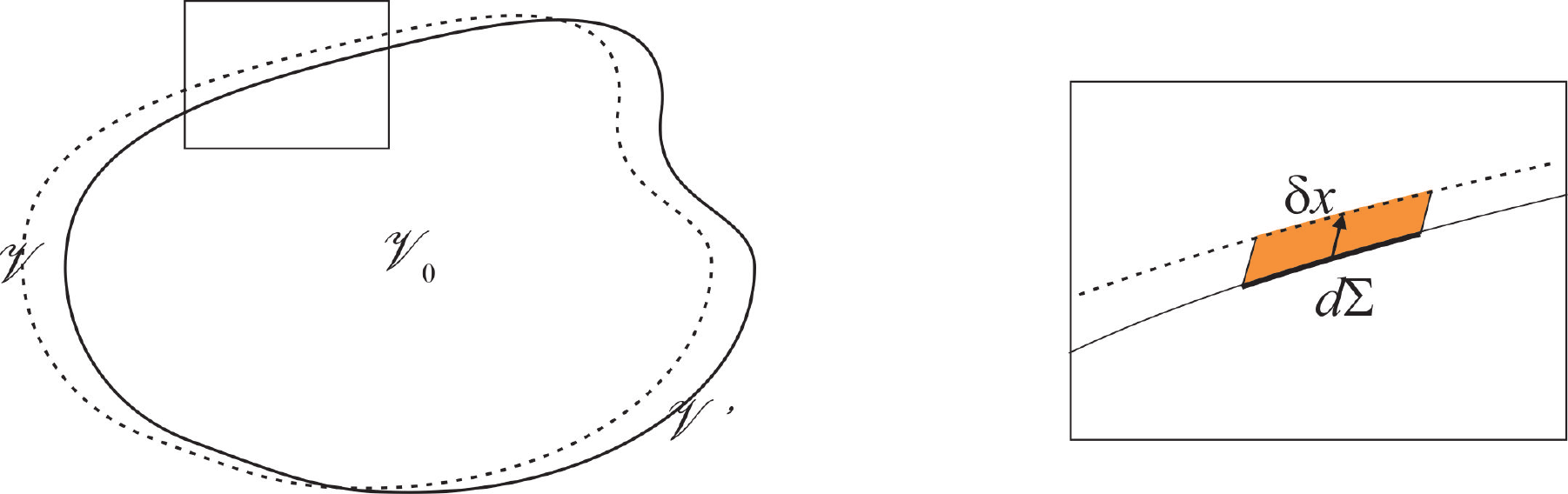}
\end{center}
\caption{The variation of the spacetime domain $\mathcal{V}$}
\end{figure}
The volume element near the boundary can be written as 
$\quad \D^4 x = \D\Sigma_a\,\delta x^a \,$, where $\,\D\Sigma_a\,$ is the hypersurface element on the boundary.
Hence equation (\ref{A4}) becomes
\begin{equation}  \label{A5}
 \Delta S(\mathcal{V}) = \int_{\mathcal{V}} \left[ \mathcal{L}\left([\phi^{\prime \alpha}],x\right) - \mathcal{L}\left([\phi^\alpha],x\right)\right]\,\D^4 x + \int_{\partial \mathcal{V}} \mathcal{L}\,\delta x^a \,\D\Sigma_a  \,,
\end{equation}
where the variation of the Lagrangian density is
$$ \mathcal{L}\left([\phi^{\prime \alpha}],x\right) - \mathcal{L}\left([\phi^\alpha],x\right) = 
\int_{\mathbb{R}^4} \D y\,\Lambda_\alpha(x,y)\,\delta \phi^\alpha(y)  \,,$$
with $\,\Lambda_\alpha(x,y)\,$ defined in (\ref{A2}).

Using this, introducing the variable $\xi= y -x\,$ and applying Gauss theorem, equation (\ref{A5}) becomes
$$ \Delta S(\mathcal{V}) = \int_{\mathcal{V}}\D x\, \left[ \partial_b\left(\mathcal{L}\,\delta x^b\right) + \int_{\mathbb{R}^4} \D\xi  \,\Lambda_\alpha(x,x+\xi)\,\delta\phi^\alpha(x+\xi) \right] $$
and, including (\ref{A2}), we arrive at
\begin{eqnarray} 
\lefteqn{ \Delta S(\mathcal{V}) -  \int_{\mathcal{V}}\D x\,\Pi_\alpha(x) \,\delta\phi^\alpha(x) = 
\int_{\mathcal{V}}\D x\, \left\{\partial_b\left(\mathcal{L}\,\delta x^b\right) + \right.} \nonumber \\[2ex]   \label{A6}
 & & \left. \hspace*{4em}\int_{\mathbb{R}^4} \D\xi  \left[ \Lambda_\alpha(x,x+\xi)\,\delta\phi^\alpha(x+\xi) - \Lambda_\alpha(x-\xi,x)\,\delta\phi^\alpha(x)   \right]\right\}
\end{eqnarray}

Now we use the identity
\begin{eqnarray*}
\lefteqn{\Lambda_\alpha(x,x+\xi)\,\delta\phi^\alpha(x+\xi) - \Lambda_\alpha(x-\xi,x)\,\delta\phi^\alpha(x) =} \\[1ex]
 &  =& \int_0^1 \D\lambda \,\frac{\D\;}{\D\lambda}\left\{\Lambda_\alpha(x+[\lambda-1]\xi,x+\lambda\xi)\,\delta\phi^\alpha(x+\lambda\xi) \right\}
\\[1ex]
 & =& \int_0^1 \D\lambda \, \xi^b \partial_b \left\{\Lambda_\alpha(x+[\lambda-1]\xi,x+\lambda\xi)\,\delta\phi^\alpha(x+\lambda\xi) \right]\} 
\end{eqnarray*}
which, combined with (\ref{A6}), leads to
\begin{equation}  \label{A7}
 \Delta S(\mathcal{V}) + \int_{\mathcal{V}}\D x\,\left[\partial_b J^b -\Pi_\alpha(x) \,\delta\phi^\alpha(x) \right] = 0 \,,
\end{equation}
where
\begin{equation}  \label{A8}
J^b = - \mathcal{L}\,\delta x^b - \int_{\mathbb{R}^4} \D\xi \, \xi^b \int_0^1 \D\lambda \,\Lambda_\alpha(x+[\lambda-1]\xi,x+\lambda\xi)\,\delta\phi^\alpha(x+\lambda\xi)  
\end{equation}

Then the local conservation of the current (\ref{A8})  [Noether theorem] 
\begin{equation}  \label{A8a}
\partial_b J^b = 0
\end{equation}
follows from the identity (\ref{A7}), provided that the Lagrangian is invariant under the transformation (\ref{A3}) and $\phi^\alpha$ is a solution of the Euler equations (\ref{A2}).

\subsection*{Maxwell field in dispersive media}
Let us apply the above results to the Lagrangian (\ref{e20})
\begin{equation}  \label{A9}
\mathcal{L} =  \frac12\, A_{b;a}(x) \,H^{ab}(x) \,, \qquad {\rm where} \qquad \,H^{ab}= 2\,\tilde{M}^{abcd}\ast A_{d;c}  \,.
\end{equation}
(the field that we had generically written as $\phi^\alpha$ has been replaced by $A_b$).

\subsubsection*{The field equations.} 
The functional derivative (\ref{A2}) is 
\begin{equation}  \label{A10}
 \Lambda^b(x,y)=  \frac12\,\delta_{;a}(x-y) \,H^{ab}(x) +  A_{e;a}(x)\, \tilde{M}^{aecb}_{\quad;c}(x-y) \,,
\end{equation}
where 
$$ \frac{\delta A_{e;a}(x)}{\delta A_b(y)}= \delta_e^b\,\delta_{;a}(x-y) \qquad {\rm and} \qquad 
 \frac{\delta H^{ae}(x)}{\delta A_b(y)} = 2 \,\tilde{M}^{aecb}_{\quad\;\, ;c}(x-y) $$
have been included. As a consequence, the Euler equation (\ref{A2}) is
\begin{equation}  \label{A11}
\Pi^b(y) \equiv - \partial_a H^{ab}(y) = 0
\end{equation}

\subsubsection*{Poincar\'e transformations. Noether theorem. }
Using (\ref{A10}) and the symmetry condition (\ref{e19a}),  the integrand of (\ref{A8}) becomes
\begin{eqnarray*} 
\lefteqn{\Lambda^f(x+[\lambda-1]\xi,x+\lambda\xi)\,\delta A_f(x+\lambda\xi) =} \\[1ex] 
 & \qquad & -\frac12\,\left\{\delta_{;a}(\xi)\,H^{af}(x+[\lambda-1]\xi) + F_{ae}(x+[\lambda-1]\xi)\,\tilde{M}^{cfae}_{\quad\;\, ;c}(\xi)\right\}\,\delta A_f(x+\lambda\xi) 
\end{eqnarray*}
and therefore
\begin{equation}  \label{A12}
J^b = - \mathcal{L}\,\delta x^b - \frac12\,H^{bf}(x)\,\delta A_f(x) + \frac12\,\int_{\mathbb{R}^4} \D\xi\,\xi^b \int_0^1\D\lambda\,
F_{ae}(x+[\lambda-1]\xi)\,\tilde{M}^{cfae}_{\quad\; ;c}(\xi)\,\delta A_f(x+\lambda\xi)
\end{equation}

Infinitesimal Poincar\'e  transformations act on coordinates as 
\begin{equation}  \label{N1}
x^{\prime a} = x^a + \delta x^a\,, \qquad \delta x^a = \varepsilon^a + \omega^a_{\;b} x^b\,, \qquad \omega_{ab}+\omega_{ba}=0 \,,
\end{equation}
where $\,\varepsilon^a\,$ and $\,\omega^a_{\;b}\,$ are constants and $\; \omega_{ab} = \eta_{ac}\omega^c_{\;b} \,$. On its turn the field $A_a$ transforms as a covariant vector, 
\begin{equation}  \label{N2}
A^\prime_a(x^\prime) = A_a(x) - \omega^b_{\; a} A_b(x)\,, 
\end{equation}
and therefore
\begin{equation}  \label{N3}
\delta A_a(x) := A_a^{\prime}(x) - A_a(x) = - \omega^b_{\; a} A_b(x) - A_{a;c}(x) \delta x^c \,.
\end{equation}
where  (\ref{A3}) and (\ref{A5}) have been included.

Then, using  (\ref{N1}) and (\ref{N3}), we easily obtain that the current  (\ref{A12}) can be written as
\begin{equation}  \label{N4}
 J^b = \varepsilon^a\,\mathcal{T}^{\; b}_a + \frac12\, \omega^{ac} \mathcal{J}^{\; \,\;b}_{ac} \,,
\end{equation}
where
\begin{equation}  \label{N5}
 \mathcal{T}^{\; b}_a = -\mathcal{L}\,\delta^b_a + \frac12\,H^{bc}\,A_{c;a} - \frac12\,\int_{\mathbb{R}^4} \D\xi\,\xi^b \int_0^1\D\lambda\,
F_{ce}(x+[\lambda-1]\xi)\,\tilde{M}^{dfce}_{\quad\; ;d}(\xi)\,A_{f;a}(x+\lambda\xi) \,,
\end{equation}
\begin{equation}  \label{N6}
  \mathcal{J}^{\; \,\;b}_{ac} =  2\, x_{[c}\mathcal{T}^{\; b}_{a]}  +  \mathcal{S}^{\;\,\;b}_{ac} 
\end{equation}
and 
\begin{eqnarray}  
\mathcal{S}^{\; \,\;b}_{ac} &=& H^b_{\;[c}A_{a]} + \int_{\mathbb{R}^4} \D\xi\,\xi^b\,\int_0^1\D\lambda\,F_{fe}(x+[\lambda-1]\xi)\,\tilde{M}^{dnfe}_{\quad\; ;d}(\xi)\,\times \nonumber  \\[1ex]  \label{N7}
	 & & \hspace*{7em}\left\{\,\eta_{n[a} A_{c]}(x+\lambda\xi) + \lambda \xi_{[a} A_{n;c]}(x+\lambda\xi)  	\right\}
\end{eqnarray}

In the non-dispersive case $\tilde{M}^{dnfe}(\xi) \propto \delta(\xi)$ and equations (\ref{N5}) and (\ref{N7}) become
\begin{equation}  \label{N7z}
\mathcal{T}^{\; b}_a = -\mathcal{L}\,\delta^b_a + H^{bc}\,A_{c;a}  \qquad {\rm and} \qquad \mathcal{S}^{\; \,\;b}_{ac} = 2 H^b_{\;[c}A_{a]} 
\end{equation}

\subsubsection*{A3: The Belinfante-Rosenfeld tensor }
Provided that the Lagrangian is Poincar\'e invariant, the current (\ref{N4}) is conserved and, as the infinitesimal parameters $\varepsilon^a$ and $\omega^{ac}$ are independent of each other, both components are separately conserved, 
$$ \partial_b \mathcal{T}^{\; b}_a = 0 \qquad {\rm and} \qquad  \partial_b \mathcal{J}^{\; \,\;b}_{ac} = 0 $$
which, including (\ref{N6}), amount to
\begin{equation}  \label{N8}
  \partial_b \mathcal{T}^{\; b}_a = 0 \qquad {\rm and} \qquad  \mathcal{T}_{[ac]} = -\frac12\, \partial_b \mathcal{S}^{\;\,\;b}_{ac} 
\end{equation}
As a consequence, if the Lagrangian is Poincar\'e invariant and the spin current vanishes, $\,\mathcal{T}_{ac}\,$ is symmetric, otherwise it need not be.

In the case of our Lagrangian (\ref{A9}), the energy-momentum tensor $ \mathcal{T}^{ab}$ (\ref{N5}) is not symmetric; however there is a technique ---see e.g. \cite{Dixon} and \cite{Landau} to quote a few---that allows to construct the Belinfante-Rosenfeld energy-momentum tensor $ \Theta^{ca}$ which ---provided that $\mathcal{L}$ is Poincar\'e invariant--- is symmetric and in some sense ``equivalent'' to $ \mathcal{T}^{ca}$ because
\begin{description}
\item[(a)] the total energy-momentum contained in a hyperplane $t= \,$constant is the same for both tensors
$$ \int \D\mathbf{x}\, \Theta^{\;4}_b (\mathbf{x},t)=   \int \D\mathbf{x}\, \mathcal{T}^{\;4}_b (\mathbf{x},t)\,$$
\item[(b)] the 4-divergences are equal too, $\quad \partial_a \Theta^{\;a}_b = \partial_a\mathcal{T}^{\;a}_b = 0 \quad$ and 
the current $\,\Theta_c^{\; a}\,$ is also conserved,
\item[(c)]  and the new orbital angular momentum current $\,2\,x_{[a} \Theta_{c]}^{\;b} \, $ and the new spin current \\
$\quad \Sigma^{\;\, \;b}_{ca}= \mathcal{J}^{\;\, \;b}_{ca} - 2\,x_{[a} \Theta_{c]}^{\;\,b} \, $ are separately conserved.
\end{description}
This is achieved by defining
\begin{equation}  \label{N9}
\Theta^{ca} = \mathcal{T}^{ca} + \partial_b \mathcal{W}^{bac} \,,\qquad \quad {\rm where}
\qquad \quad  \mathcal{W}^{bac} = - \mathcal{W}^{abc}
\end{equation}
and
\begin{equation}  \label{N10}
\mathcal{W}^{bac}  = \frac12\,\left(\mathcal{S}^{bac} + \mathcal{S}^{bca} - \mathcal{S}^{acb}  \right) \,, 
\end{equation}
To avoid ambiguity one usually refers to $\mathcal{T}^{ca}$ as the {\em canonical} energy-momentum tensor and to $\Theta^{ca}$ as the {\em Belinfante-Rosenfeld} tensor. 

In our present case, substituting (\ref{N7}) in (\ref{N10}) leads to
\begin{eqnarray}
\mathcal{W}^{bac} & = & \frac12\,H^{ba} A^c + \frac12\,\int_{\mathbb{R}^4} \D\xi\,\int_0^1\D\lambda\,F_{fe}(x+[\lambda-1]\xi)\,\tilde{M}^{dnfe}_{\quad\; ;d}(\xi)\,\left\{2 \lambda\,\xi^c \,\xi^{[b} A_{;n}^{\;\;a]}(x+\lambda\xi)  \right. \nonumber  \\[1ex]  \label{N11}
	 & & \left. \hspace*{9em} + \left(\delta_n^{[b} \eta^{a]m} \xi^c + \delta_n^{[b} \eta^{c]m} \xi^a  - \delta_n^{[a} \eta^{c]m} \xi^b      \right)\,A_m(x+\lambda\xi) \right\}
\end{eqnarray}
When calculating $ \,\partial_b \mathcal{W}^{bac}\,$ the combination $\xi^b\partial_b$ will occur in several instances like
$$ \xi^b\partial_b \Psi(x+\lambda\xi,\xi) = \frac{\partial\;}{\partial \lambda}\,\Psi(x+\lambda\xi,\xi) \,,$$
where $ \Psi(x+\lambda\xi,\xi) $ is a product of $F_{fe}(x+\lambda\xi -\xi)$ times either $A_m(x+\lambda\xi)$ or its derivative. This permits to perform some integrals on $\lambda$ like
\begin{eqnarray*}
\int_0^1\D\lambda\,\xi^b\partial_b \Psi(x+\lambda\xi,\xi) &=& \Psi(x,x+\xi) - \Psi(x-\xi,x) \\[1ex]
\int_0^1\D\lambda\,\lambda \,\xi^b\partial_b \Psi(x+\lambda\xi,\xi) &=&  \Psi(x,x+\xi) - \int_0^1\D\lambda \,\Psi(x+\lambda\xi,\xi)  
\end{eqnarray*}
Hence, from (\ref{N11}) and after a little algebra, it follows that
\begin{eqnarray}
\partial_b\mathcal{W}^{bac} & = & \frac12\,H^{ba} A^c_{\; ;b} + \frac12\,F_{fe}\,\int_{\mathbb{R}^4} \D\xi\,\tilde{M}^{dnfe}_{\quad\; ;d}(\xi) \left[\xi^c\,A_n^{\; ;a}(x+\xi) + \delta_n^{[c} A^{a]}(x+\xi)   \right] \nonumber\\[2ex]  
 & & + \frac12\,\int_{\mathbb{R}^4} \D\xi\,\int_0^1\D\lambda\,\tilde{M}^{dnfe}_{\quad\; ;d}(\xi)\,\times\left\{ \delta_n^{[b}\left( \eta^{a]m} \xi^c + \eta^{c]m} \xi^a \right)\, \partial_b\left[F_{fe}(X-\xi)\,A_m(X) \right] \right. \nonumber\\[2ex]  \label{N12} 
 & & \left.  - \xi^c F_{fe}(X-\xi)\,A_n^{\; ;a}(X) - \lambda \xi^c \xi^a \partial_b\left[F_{fe}(X-\xi)\,A_n^{\; ;b}(X) \right] \right\} \,, 
\end{eqnarray}
with $\, X = x+\lambda\xi\,$, which substituted in (\ref{N9}) leads to
\begin{eqnarray}
\Theta^{ba}  &=& \frac12\, H^{ca} F_c^{\;\,b} -\frac14\,\eta^{ab} F_{ed} H^{ed} + 
\frac12\,F_{fe} \left[\tilde{M}^{fed[b}\ast F^{a]}_{\;\,d} + 
\tilde{M}^{fed(b;a)} \ast A_d + \frac12 \left( y^b \tilde{M}^{fedn}\right)\ast F_{dn}^{\;\;;a} \right] \nonumber \\[2ex]
 & & - \frac12\, \int \D^4 \xi \tilde{M}^{ndfe}(\xi)\,\frac{\partial\;\,}{\partial \xi^d}\,\int_0^1 \D\lambda \,\xi^{(a}\left\{ F_{fe}(X-\xi) \left[A_n^{\;;b)}(X) + F^{b)}_{\;\,n}(X)\right] \right. \nonumber\\[2ex] \label{D4A}
 & &  - F_{fe;n}(X- \xi) A^{b)}(X)  + \left. \delta^{b)}_n \left[F_{fe}(X- y) A^c(X) \right]_{;c} + \lambda  \xi^{b)} \left[F_{fe}(X- \xi) A_n^{\;;c}(X)  \right]_{;c} \right\}
\end{eqnarray}
where $X = x + \lambda \xi\,$.
For non-dispersive media $\tilde{M}^{ndfe}$ is constant and both tensors, $\mathcal{T}^{ba}$ and $\Theta^{ba}$, reduce to the already known (\ref{e17a}) and (\ref{e17d}), respectively.


\end{document}